\pdfoutput=1 
\documentclass{JINST}

\title{A Specialized Processor for Track Reconstruction
at the LHC Crossing Rate}

\author{A.~Abba$^b$, F.~Bedeschi$^c$, M.~Citterio$^b$, F.~Caponio$^b$, A.~Cusimano$^b$, A.~Geraci$^b$, P.~Marino$^c$, M.J.~Morello$^c$, N.~Neri$^b$, G.~Punzi$^c$\thanks{Corresponding author.}, A.~Piucci$^c$, L.~Ristori$^{cd}$, F.~Spinella$^c$, S.~Stracka$^c$, D.~Tonelli$^a$\\
\llap{$^a$} CERN,\\
  Geneva, Switzerland\\
\llap{$^b$} Politecnico and INFN-Milano,\\
  Milano, Italy\\
\llap{$^c$}University, SNS, and INFN-Pisa,\\
  L.go Pontecorvo 3, 56127, Pisa, Italy\\
\llap{$^d $}Fermilab\\
  Batavia, IL, USA\\
E-mail: \email{giovanni.punzi@pi.infn.it} }

\abstract{We present the results of an R\&D study of a specialized processor capable of precisely reconstructing events with hundreds of charged-particle tracks in pixel detectors at 40 MHz, thus suitable for processing LHC events at the full crossing frequency.  For this purpose we design and test a massively parallel pattern-recognition algorithm, inspired by studies of the processing of visual images by the brain as it happens in nature. We find that high-quality tracking in large detectors is possible with sub-$\mu$s latencies when this algorithm is implemented in modern, high-speed, high-bandwidth FPGA devices. This opens a possibility of making track reconstruction happen transparently as part of the detector readout.}


\begin{document}

\section{Introduction}
\label{sec:Introduction}

In modern experiments at high-energy hadron colliders, data acquisition and reconstruction pose great technological challenges. Tracking systems play a central role in this issue, due to the large combinatorial and the size of the associated information flow. Tracks are needed to reconstruct and quickly select potentially interesting events for higher level of processing, and finally permanent storage for subsequent analysis. 

A great deal of study has been devoted to the fast reconstruction of tracks, with specialized processors developed explicitly for this purpose. The difficulty of the problem calls for a high level of parallelization. A notable example of specialized processors developed for real-time tracking is the Associative Memory~\cite{Dell'Orso:1988zz}. The fast parallel matching capability of the AM, in conjunction with a second-stage process of lineared fitting, has allowed tracks to be reconstructed in complex detectors at frequencies of 30-100 kHz, and with latencies of order 10-20$\mu$s, leading to a track-based trigger to be implemented in CDF at the second trigger level. The FTK processor, based on similar principles, is currently being incorporated in the second level of the trigger in the ATLAS experiment~\cite{FTK}.

However, the complexity of events at LHC, and even more the future High-Luminosity LHC, calls for even greater performance. In the more complex environment of HL-LHC the multiplicity of collisions per crossing increases dramatically, and global, simple variables become inadequate for an effective selection of events. This affects flavor-physics measurements most severely, but extends to other areas as well, like jet reconstruction and general event topology. As a consequence, the role of tracking is further enhanced, and calls for track reconstruction at even earlier stage of the data acquisition chain.  

All of this means that a device capable of reconstructing tracks at the full LHC crossing rate of 40 MHz would be highly desirable.
This however requires a large jump forward with respect to current systems, that cannot simply be obtained on the basis of industry's progress in gate density and speed. Even at clock frequencies of 1 GHz, one would need a device capable of a complex reconstruction in just $\simeq$25 clock cycles. For comparison, if one normalizes the throughput of currently existing AM-based devices to their clock frequency, one finds that tracking of an event takes of the order of 1000-2000 clock cycles (unspecialized CPU-like architectures typically require about $10^8$ cycles per event).

It may seem hard to fill such a large gap, and there are indeed no known examples of artificial devices with this kind of performance. An unconventional example, however, may be found in the natural world. It is known, for instance, that the early visual areas in human brain cortex are able to extract the main feature of visual images at rates of $\simeq$30Hz. While this may seem a low rate by HEP standards, when compared with the typical firing frequency of brain neurons of $\simeq$1kHz, it corresponds to $\simeq$30 time units per image, which is the ballpark performance we would need to achieve at LHC. While natural vision seems, and in many ways is, a different kind of process, experimental studies have shown that the essential features of the early vision processing are, at a detailed functional level, quite similar to a process of fast track reconstruction\cite{PDV}. This gives some hope that this kind of performance might be attainable for a HEP tracker processor as well, if the right algorithms are used.  An important features behind this fast performance seem to be a fully parallelized pattern-matching process, which is "analog", rather than digital. This allows averaging over nearby elements to extract a significant better resolution than the available cell granularity, in a single, fast processing step, rather than as a two-step procedure typical of AM-based systems.
Studies along this direction have already been performed in the past. An algorithm named {\it artificial retina} has been developed based on these principles, and applied to the reconstruction of 2D straight tracks in an idealized detector\cite{Ristori:2000vg}. The basic idea of mapping the response of several hits onto a fixed array of cells representing the parameter space of tracks is present also in older studies ({\em Hough transform}~\cite{Hough:1959qva}), although based on the simpler technique of a "voting" procedure for a cell in the parameter space, rather than the more refined, biologically-inspired mechanism of weighted response.

In the present work, we have developed for the first time a concrete implementation of these general principles to 3D real-time tracking in a realistic LHC detector under realistic operating conditions, using electronic components that are commercially available on the market at the present time. 
We chose to focus our work on a section of the LHCb upgrade detector, where real-time tracking plays a particularly important role.

To achieve our goals, we had to introduce further biologically-inspired elements in the design, the most important of them being the exploitation of a strong connectivity, spreading the workload over an extended, high-bandwidth network of elements, thus allowing a greater degree of parallelization and a better exploitation of resources.

\section{Simulation}
\subsection{Principles of the {\it artificial retina} algorithm}

We briefly recall here the basic principles of the algorithm.

Consider a single straight track traversing an array of $n$ parallel detector layers. The parameter space of all possible tracks is divided into a grid of receptive fields, the {\it cells}. The metric and dimensionality of this space depends on the problem at hand, but we will use a simple 2D example for illustration, with two parameters $(p,q)$. Each parameter-space cell is uniquely associated with a set of coordinates $(p_i, q_j)$. These coordinates correspond to the {\it intersections} that a track with parameters identified by the center of the cell has in the measurement planes. 
For each incoming hit, the algorithm computes the excitation intensity of the cell corresponding to $(p_i, q_j)$
\begin{equation}
R_{ij} = \Sigma_{k=1, r}^{k=n} \exp \left ( - s^2_{ijkr}/2\sigma^2\right)
\end{equation}
using the distance 
\begin{equation}
s_{ijkr} = x^{(k)}_{r}  - y_k(p_i, q_j)
\end{equation}
between the hit position $x^{(k)}_{r}$ and the intersection of the $(p_i, q_j)$ track $y_k(p_i, q_j)$ on layer $k$. The sum runs over all hits in all layers and the excitation $R_{ij}$ is computed for all cells. In this example a Gaussian excitation function is adopted for illustration, but other functions can be used. The width parameter $\sigma$ can be adjusted to optimize the sharpness of the response of the receptors.

 After all hits are processed, tracks are identified as local maxima in the cell space, via a local cluster-finding algorithm.

\begin{figure}[tbp]
\begin{center}
\includegraphics[width=0.9\textwidth]{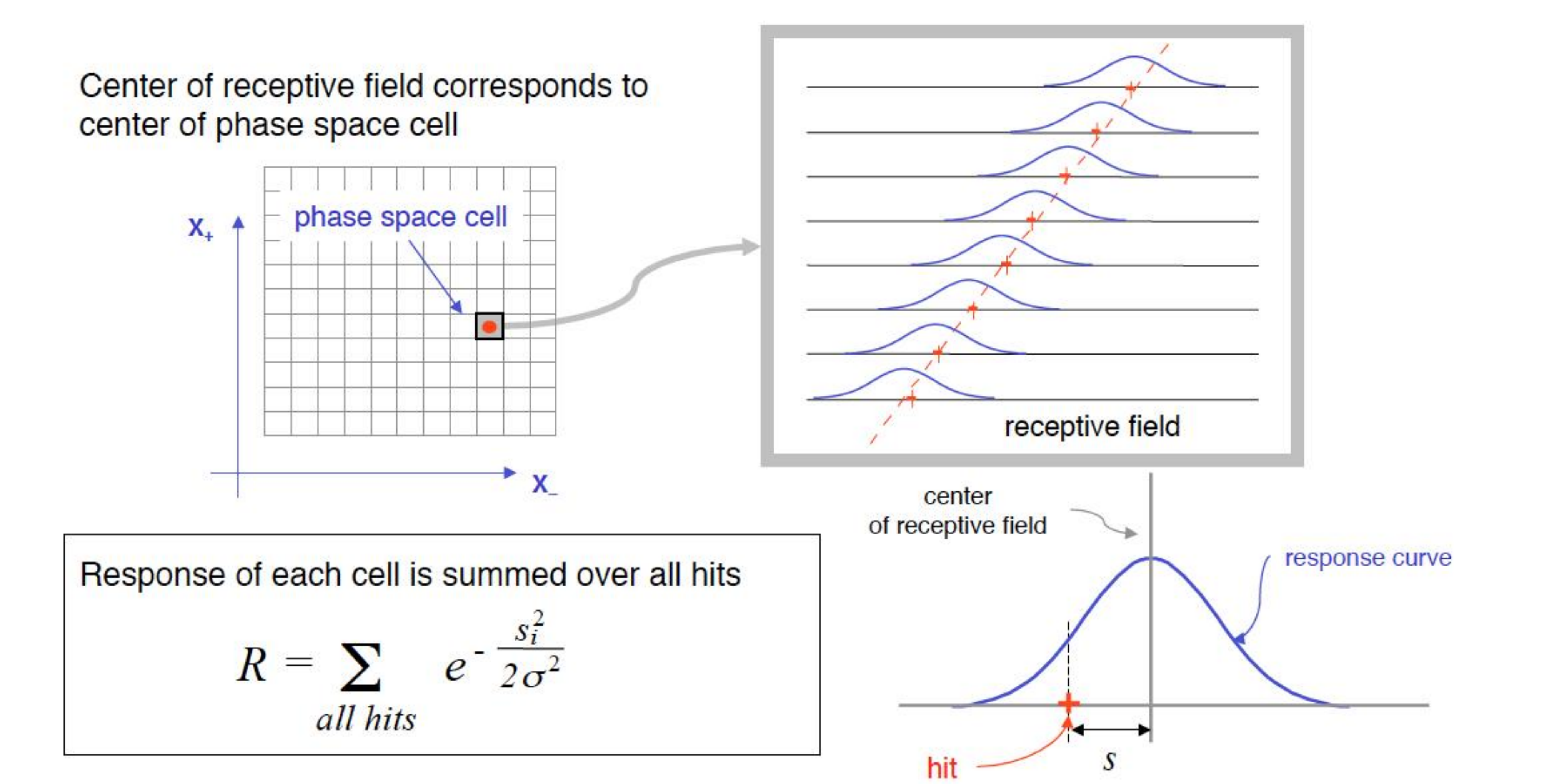}
\end{center}
\caption{Illustration of the retina-algorithm concept. (Reproduced with permission from Ref.~\cite{luciano-instr99}.)}
\label{fig:retina}
\end{figure}

\subsection{Tracker Geometry}
\label{sec:geometry}

\begin{figure}[tbp]
\begin{center}

\includegraphics[width=0.6\textwidth]{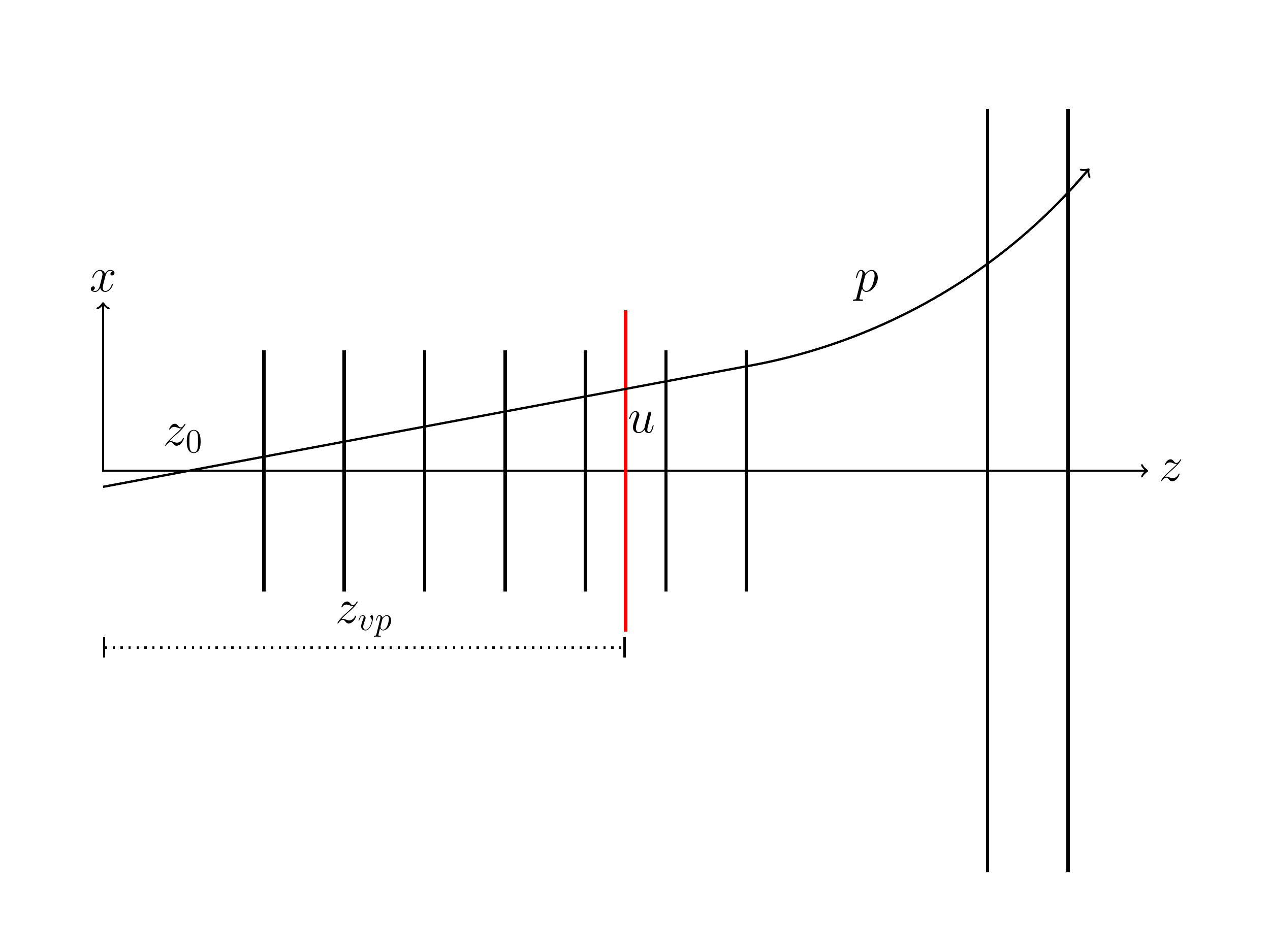}

\end{center}
\caption{Illustration of track parameterization.}
\label{fig:track_parameters}
\end{figure}

We consider an array of tracking detectors in a generic magnetic field. The 3D trajectory of a charged particle originating outside the tracker can be uniquely identified by five parameters, independently of the uniformity of the magnetic field or the alignment of the detectors. For instance, one might consider the 3 coordinates of the originating vertex, plus the 3 components of the particle's 3-momentum - minus one parameter representing the unknown physical point of origin of the particle along its trajectory.
A 5-parameter space may easily lead to an impractically large number of cells, and this may be a big hurdle for this approach to the problem. We chose the approach of selecting only 2 dimensions as "main parameters", to be counted on for pattern recognition of tracks, leaving the other parameters to the treated as "perturbations". 
For the two main coordinates, we have chosen the spatial coordinates of the intersection of the track with a plane perpendicular to the detector ($z$) axis. There is no fundamental motivation for this particular choice, except that it is intuitive and easy to visualize, and that it has the property that two tracks that are well-separated in the detector are most likely well separated also in this projection. The $z$ coordinate of this virtual layer is arbitrary as well, but for what has been said before, an intuitively good choice is to position it somewhere in the middle of the tracking layers (see red line in Fig.~\ref{fig:track_parameters}). No important features of the tracking performance of our algorithm are expected to be sensitive to the way these arbitrary choices are made. 

For the remaining three parameters, we picked some more standard definitions:
\begin{itemize}
\item  {\boldmath $d$} $-$ signed transverse impact parameter, defined as the distance of closest approach of the particle trajectory to the $z$-axis;
\item  {\boldmath $z$} $-$ the $z$-coordinate of the point of closest approach to the $z$-axis;
\item  {\boldmath $k$} $-$ signed track curvature, defined as $q/\sqrt{p_x^2 + p_z^2}$, where $q$ is the particle charge, and $p_x$ and $p_z$
are the momentum components perpendicular to leading magnetic field direction ({\boldmath $\vec{B}$} = {\boldmath $B\hat{y}$}), at the point of closest approach to the $z$-axis.
\end{itemize}

The (u,v,d,z,k) parameter space is then divided in cells, but a fine grid is used only in the (u,v) plane, while only a small number of bins are taken in the other directions ("lateral cells" - more details below).
In order to avoid the need for a separate, time-consuming track-fitting stage, a standard approach would require to subdivide the parameter space with a granularity corresponding to the desired parameter resolution (this also holds for the original Hough transform). However, the analog response of cells, graduated with the distance of the input hits from the "ideal track", makes it possible to use much larger cells.

Care must be taken, however, to use an appropriate metric in the parameter space. In most cases, the distribution of tracks will not be uniform in the parameter space, and this leads to an inefficient distribution of cells, if uniformly sampled. We have performed appropriate non-linear transformation of coordinate measured on the virtual plane, to achieve a track distribution which is uniform in the (u,v) space.

Having thus defined our cells in the parameter-space, each cell $(u,v,d,z,k)_i$ can then be uniquely associated with the set of coordinates ($(x_{ij}, y_{ij})$) of the intersections of a track with parameters corresponding to the center of that cell with each of the detector layers (labeled by the index $j$).
These calculated intersections are then used in computing the excitations of each cell of the array for every input hit, according to the algorithm described in the previous section.

\begin{figure}[tbp] 
\begin{center} 
\includegraphics[width=0.6\textwidth]{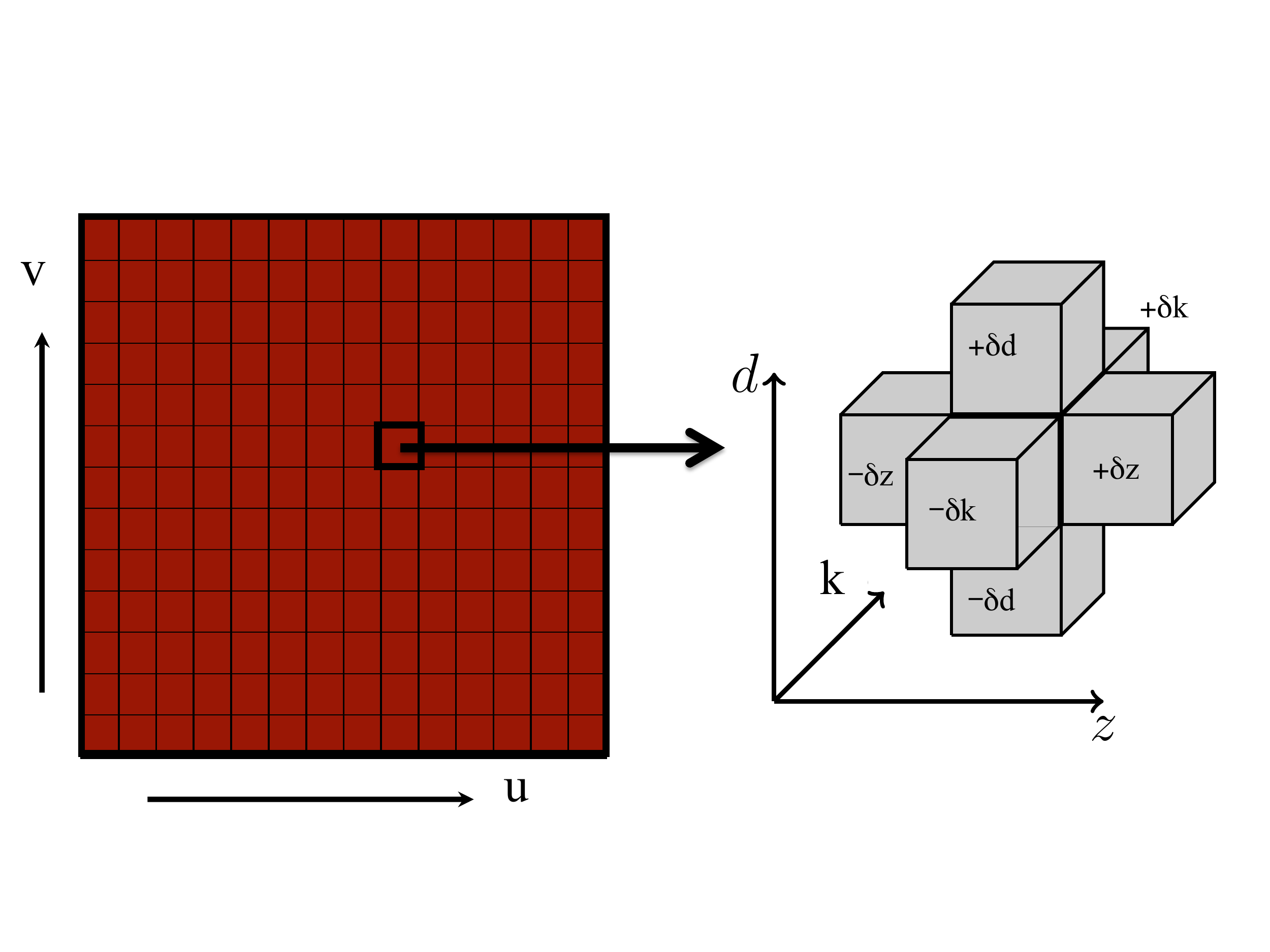}

\caption{Illustration of the main cells in the $(u,v)$ space with $d=z=k=0$ along with the lateral cells associated with compact dimensions.}
\label{fig:compact_dimension}
\end{center} 
\end{figure} 

At this level, a track is identified by a cluster over threshold in this two-dimensional space. In the second step, an estimate of track parameters is performed  by balancing the excitation found in the lateral cells for each compact dimension. For each main cell $(u,v,0,0,0)$ we fill six lateral cells
$(u,v,\pm \delta d,\pm \delta z,\pm \delta k)$, 
as shown in Fig.~\ref{fig:compact_dimension}.  The $u,v$ calculation requires finding the center of mass of a $3\times3$ square, whereas the extraction of $(d,z,k)$ requires computing the center of mass of a $3\times 3\times 3$ cube whose  only a subset of coordinates in each dimension are nonzero, thus reducing the problem to the processing of seven values.

\subsection{Results}

\begin{figure}[tbp] 
\begin{center} 
\includegraphics[width=0.6\textwidth]{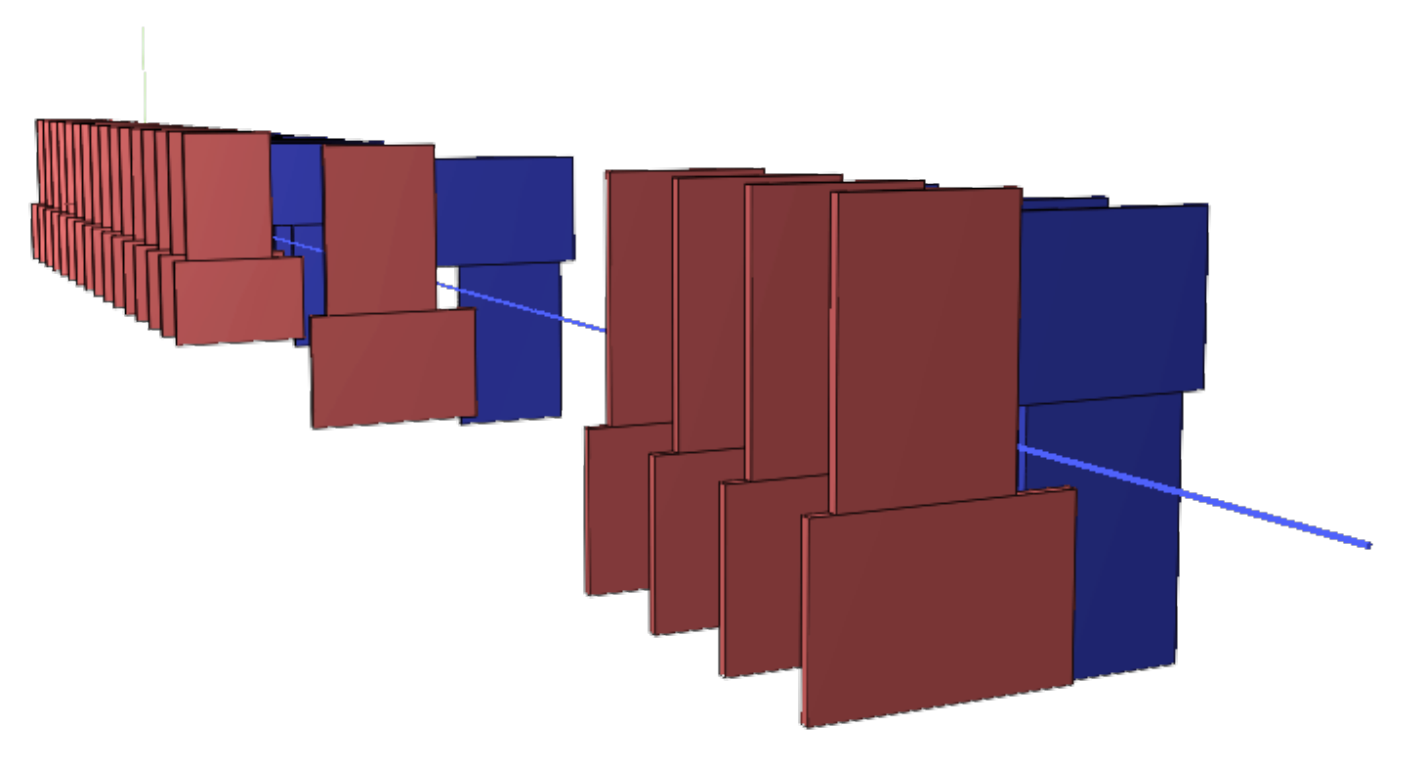}
\caption{\label{fig:VELO}3D model of the tracking detector used in our simulation.}
\end{center} 
\end{figure}

To apply this procedure to a specific example, we have picked an array of six pixel detectors, arranged according to the geometry of the upgraded LHCb VELO tracker\cite{VELOPIX} (see Fig.~\ref{fig:VELO}). The magnetic field inside this detector is negligible, so we only needed to take care of four track parameters. 
We used real LHCb data from the past run to evaluate the probability distribution of tracks in the parameter space, and derive the needed nonlinar transformations.
We discretized the main $(u,v)$ subspace into $50,000$ cells, a granularity $\mathcal{O}(100)$ larger than the maximum expected number of tracks in the event.  

\begin{figure}[tbp] 
\begin{center} 
\includegraphics[width=0.8\textwidth]{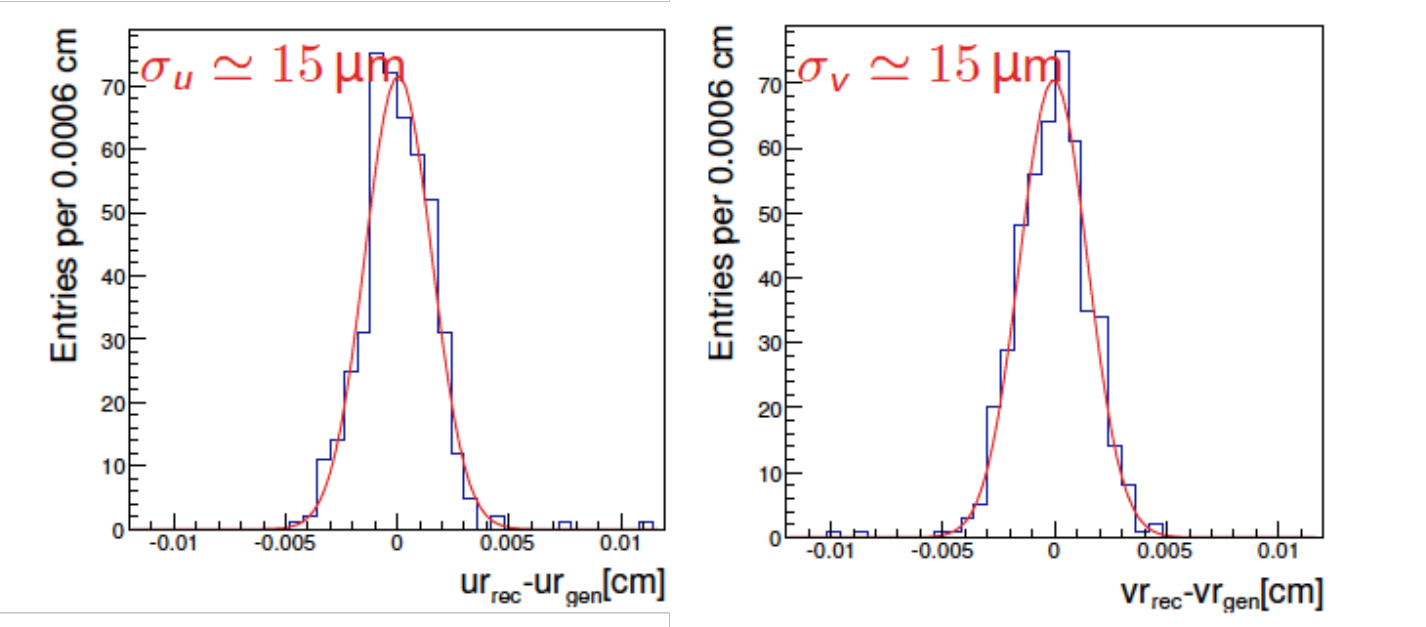}
\includegraphics[width=0.5\textwidth]{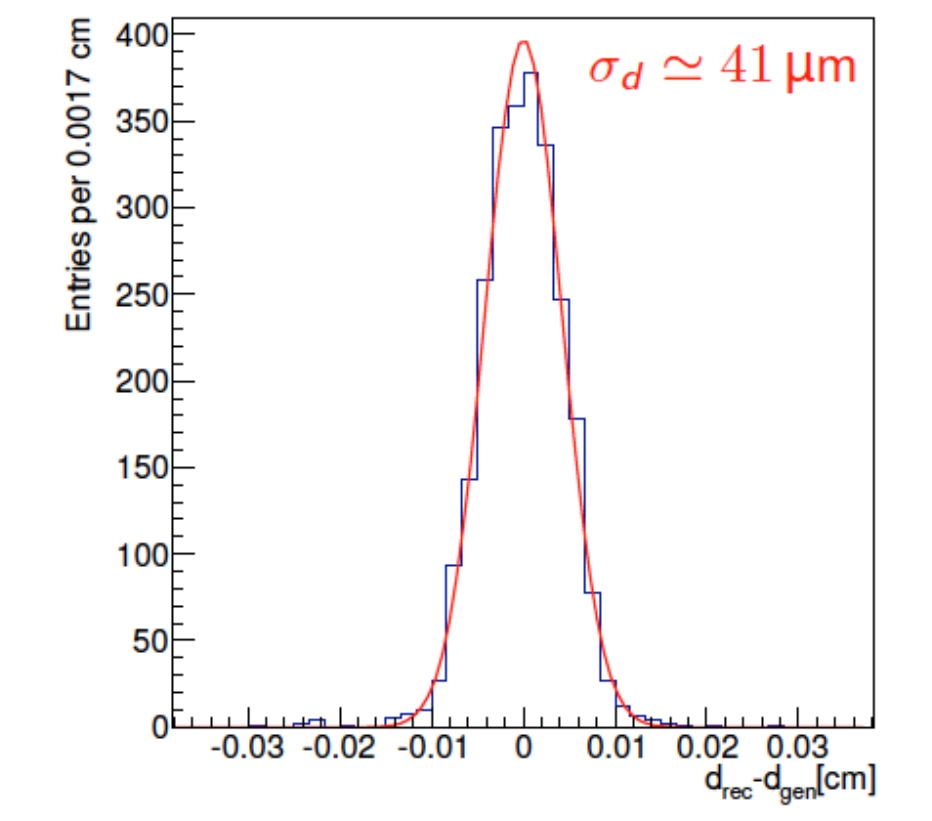}
\caption{Residuals of measured track parameters, with respect to their true values.}
\label{fig:resolutions}
\end{center} 
\end{figure}

We simulate tracks through the detector, assuming for the pixels a resolution of 12$\mu$m in both coordinates. The results, shown in Fig.~\ref{fig:resolutions} show that our system is capable of providing an excellent tracking resolution, very close to a full-fledged offline-style fit, in spite of being based on simple computations performed locally on a grid\footnote{There results were later confirmed by further, more complete simulations using the full official LHCb Monte Carlo, including all realistic detector effects and more complex detector configurations, also including magnetic field and momentum measurement. Efficiencies and ghost rates of our algorithm turned out very close to offline performances. }.

These encouraging results on the intrinsic solidity of the algorithm motivated us to pursue an hardware implementation and evaluate its performance and complexity in detail.

\begin{figure}[tbp]
\includegraphics[trim=0cm 3.5cm 0cm 0cm, clip=true,scale= 0.6]{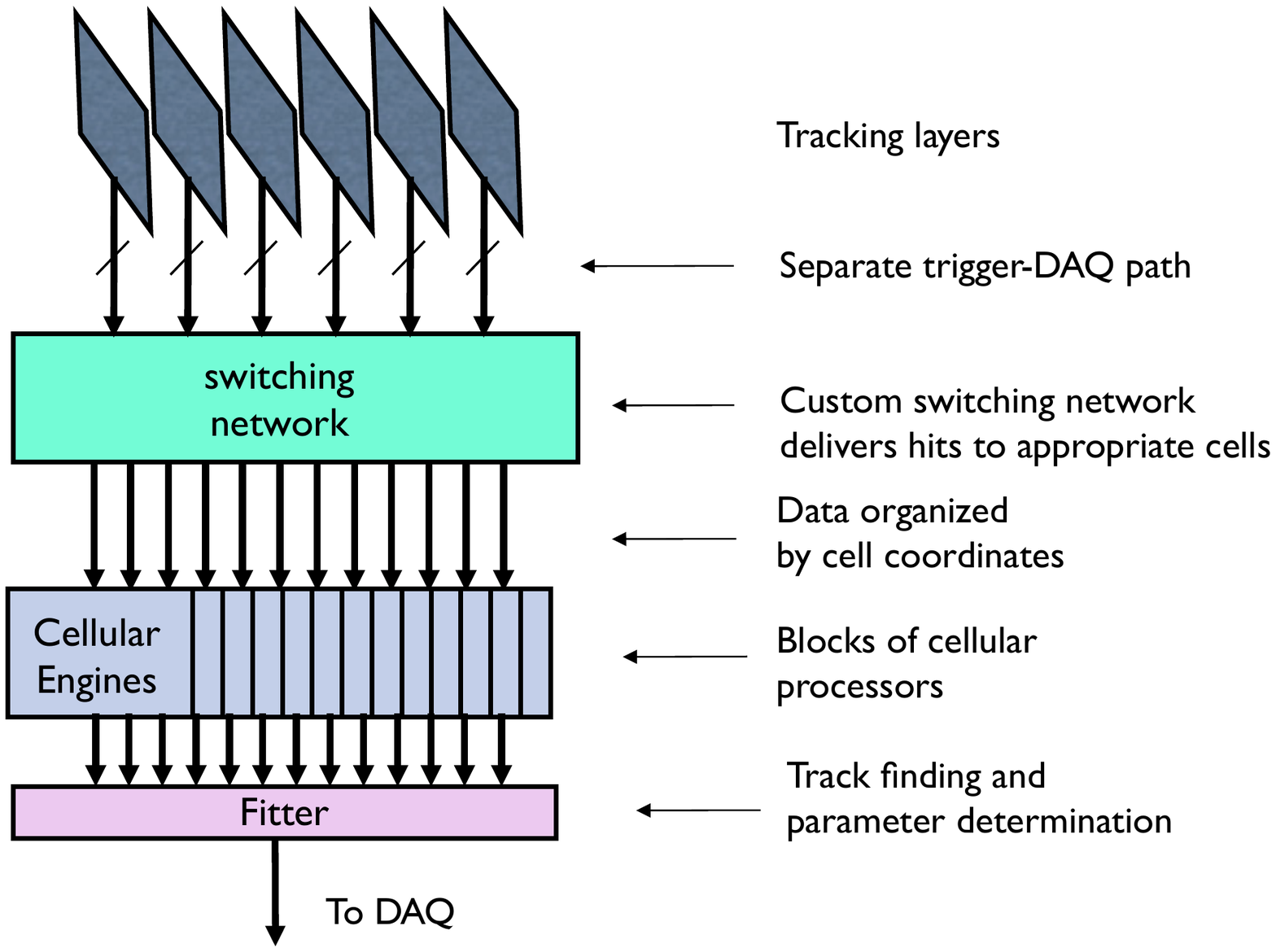}
\caption{Architecture overview.}
\label{fig:architecture}
\end{figure}

\section{Implementation}

\subsection{Architecture}

The general architecture of our proposed implementation is shown in Fig.\ref{fig:architecture}. The array of cells is mapped onto an array of cellular processors, each of them having the task of evaluating and accumulating the excitation level a single $(u,v)$ cell, together with all its associated lateral cells. Each cell is implemented as an independent block of logic ({\it processing engine}) that performs the necessary elementary operations.

A crucial issue to solve is that of fast delivering of hits to these processors. We arrange for hits to flow from the detector layers into a custom switching network devoted to this purpose. The operation of the switching network must be such as to deliver in parallel all needed hits to all the process that require them. This implies that the switch must be "smart", and contain appropriate embedded logic allowing to dispatch hits to appropriate, multiple locations, duplicating them along the way where needed.

Once the engines have finished calculating each cell excitation, local maxima must be found in parallel in all cells, via horizontal exchange of informations between adjacent processors. Then, the coordinates and intensity of the local maxima (above some threshold), and the intensities of their nearest neighbors, are output from the grid, and track parameters are extracted from the cell information.

This  architecture is flexible and linearly scalable with the number of tracking layers, the resolution, and the average detector occupancy.

\subsection{The Switch}

 This functionality of our switch is similar to that of commercial network switches, although more specialized. 

\begin{figure}[tbp]
\begin{center}
\includegraphics[width=0.45\textwidth]{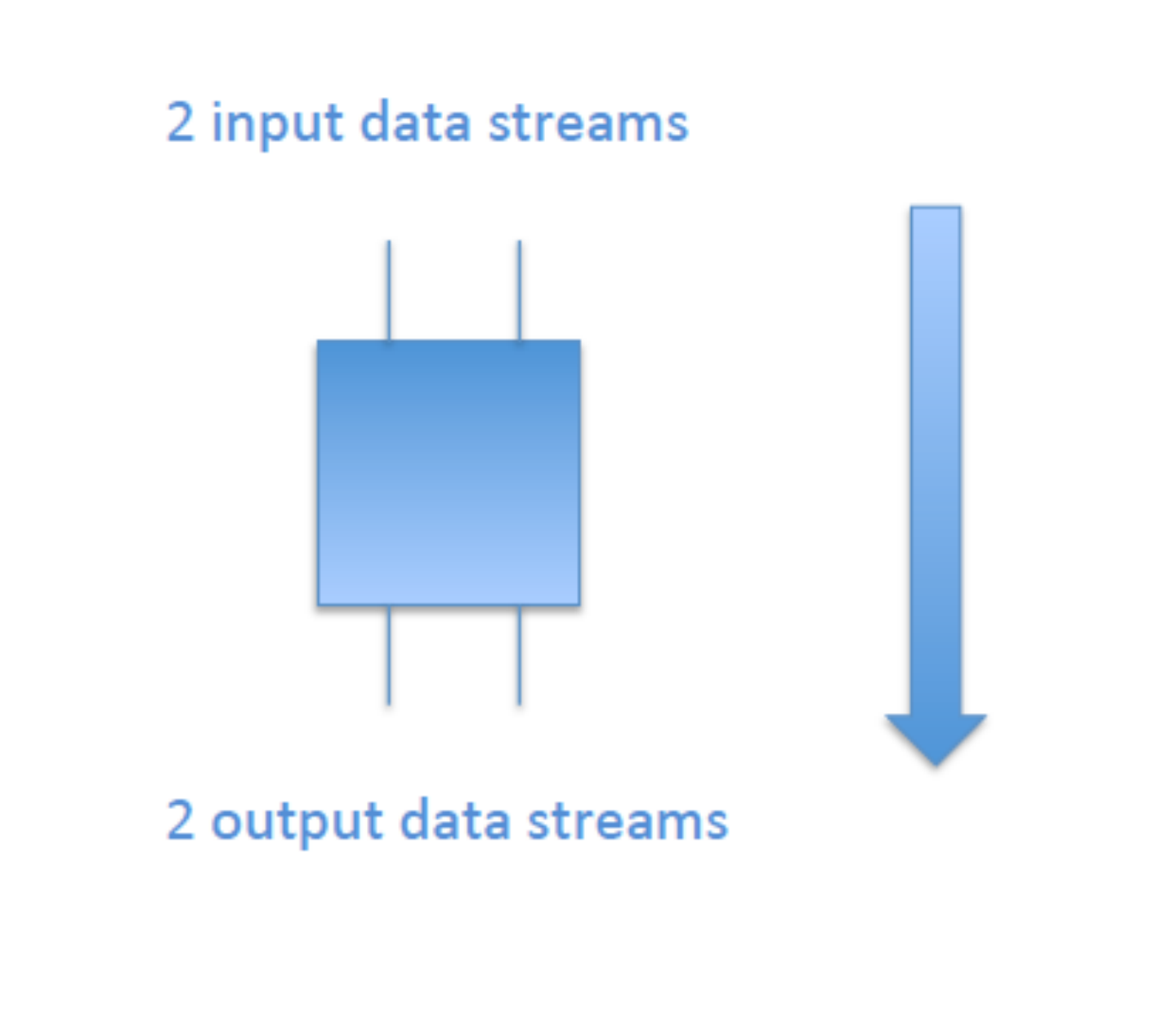}
\includegraphics[width=0.45\textwidth]{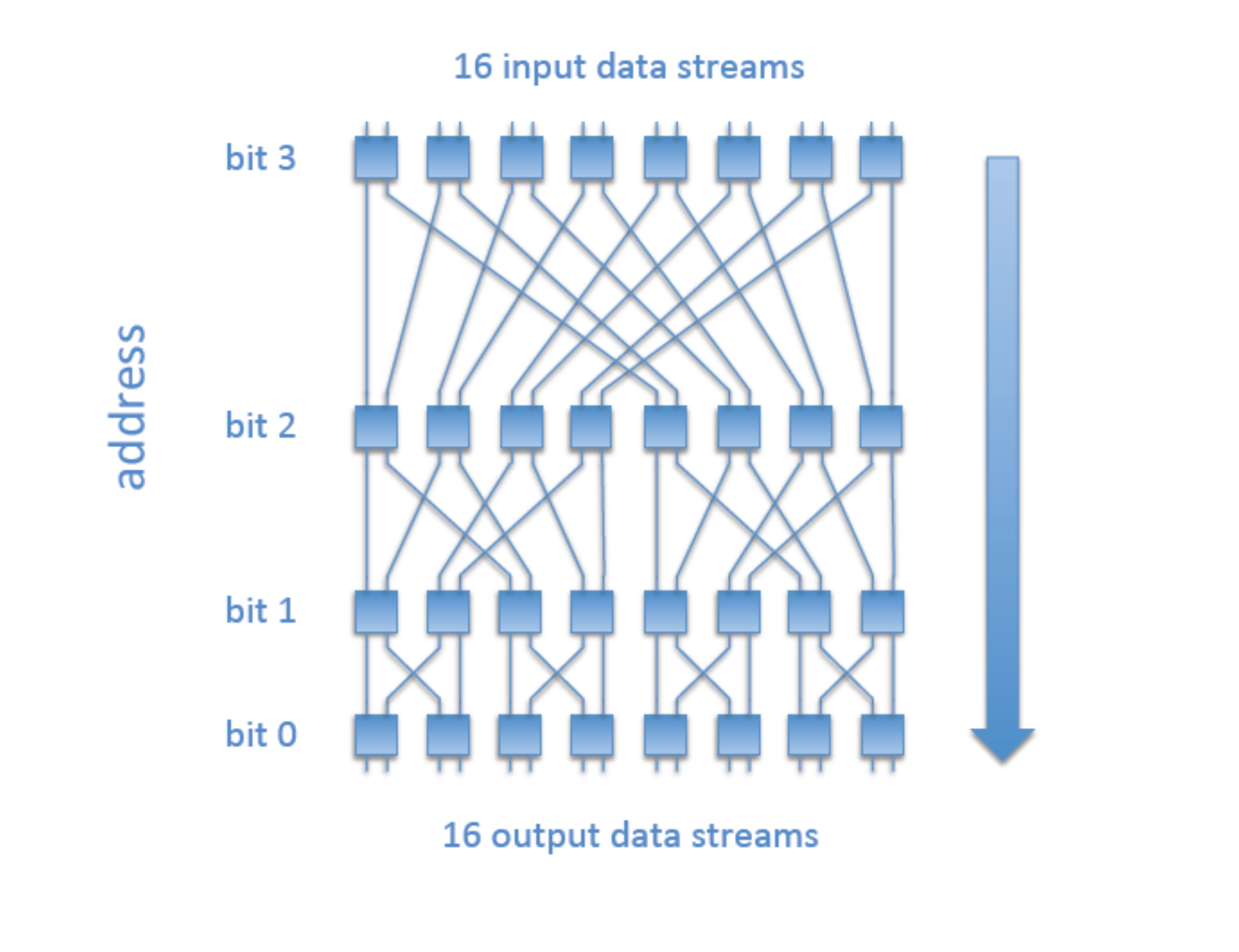}
\end{center}
\caption{Basic logic unit of the switching network.}
\label{fig:switch-concept}
\end{figure}

The switch is built from a network of nodes. The basic building blocks are two-way sorters (Fig.~\ref{fig:switch-concept}, left), with two input data and two output data streams.

A two-way sorter merges left and right input data and dispatches hits to one or both outputs according to the group the input hits belong to. Left and right inputs are merged. If a stall from downstream layers occurs, one or both input streams are held.  Such elementary building blocks are combined to build the needed network topology, with the required switching capability (Fig.~\ref{fig:switch-concept}, right). An $N \times N$ network requires $\log_2(N)N/2$ elements. The modular structure has several advantages over a monolithic design. It allows easy scalability and reconfiguration of the system when necessary, and distributing the necessary addressing information over the whole network, storing the information only at the nodes where it is required.

\subsection{The processing engine}

The engine is implemented as a clocked pipeline. 
The intersections $x_0(k)$ and $y_0(k)$ for each layer $k$ are stored in a ROM. The layer identifier associated with each incoming hit selects the appropriate set of $x_0(k)$ and $y_0(k)$ coordinates that are subtracted from the hit's $x$ and $y$ coordinates. The outcomes of the subtractions are squared, summed, and the result $R$ is rounded by keeping the eight least significant bits. The weighting function, common to all engines, is mapped into a $8 \times 256$ bit lookup table. The rounded result $R$ is used as address to the lookup table. The outputs of the lookup table are accumulated for each hit of the event. The same hit is cycled seven times in the engine logic, once for the central cell calculation, and six for the lateral cell calculations. Hence, seven accumulators are defined for each cell and one hits enters the engine every seven clock cycles. See Sec.~\ref{sec:geometry} and Fig.~\ref{fig:compact_dimension} for a description of cells configuration.

Once the readout of an event is completed, a word with the EndEvent bit arrives, prompting each engine to send the contents of its central cell to the neighboring engines. Immediately after, each engine compares the excitation in its central accumulator with the excitations received from the neighbors and raises a LookAtMe flag if it identifies its excitation as a local maximum. 
Additional logic looks at the LookAtMe flag and, if not busy, requires from the engine the content of all seven accumulators {\it and} the content of central accumulators of neighboring engines. These data are then queued out, and used to calculate the track parameters. 
 
\begin{figure}[tbp]
\begin{center}
\includegraphics[width=0.9\textwidth]{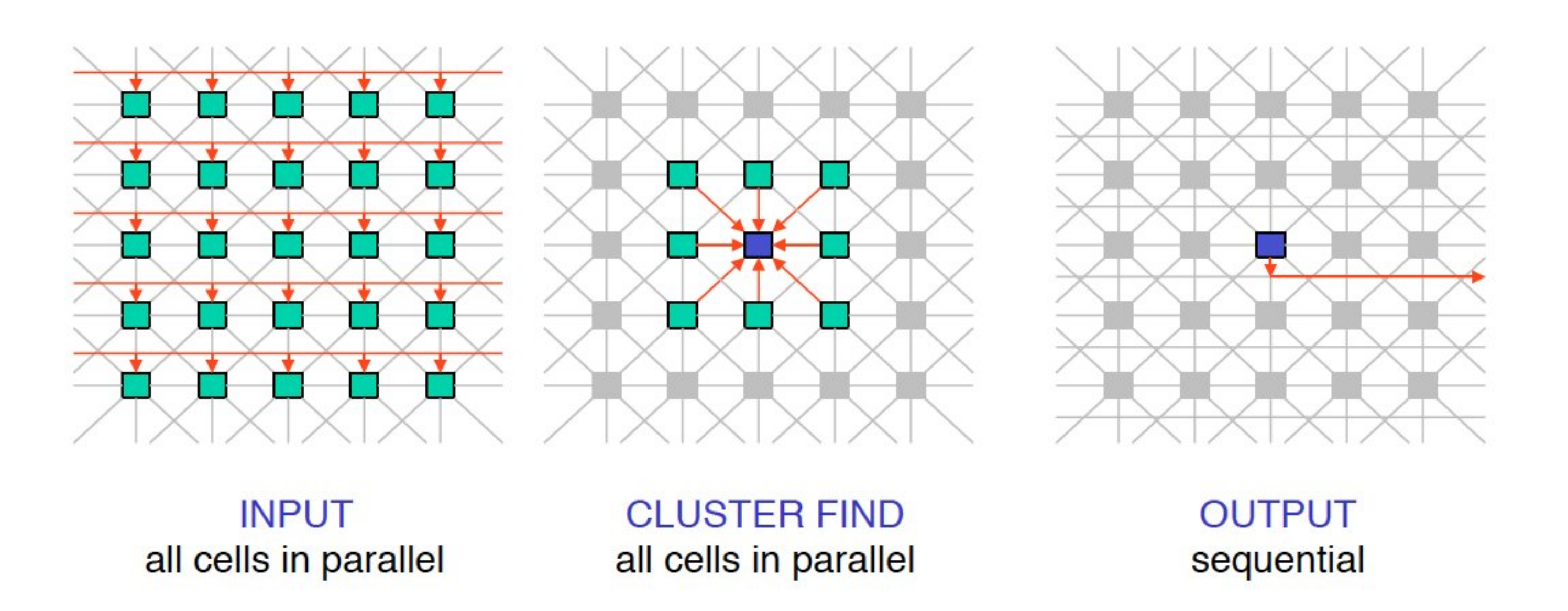}
\end{center}
\caption{Illustration of the clustering data flow.}
\label{fig:clustering}
\end{figure}

\subsection{Parameter determination}

Calculation of the cluster centroid is factorized into two separate processes. The $u,v$ calculation implies finding the center of mass of a $3\times3$ square; the $(d,p,z)$ calculations requires computing the center of mass of a $3\times 3\times 3$ cube. Since only a subset of coordinates in each dimension are nonzero, the problem reduces to the processing of seven values. The operation for each coordinate is $u =  u_0/d_k + (\sum_{i,j} u_i l_{i,j}) / \sum l$, where $u_0/d_k$ is a global translation that depends on the absolute position of the engine and is not calculated in real time, but stored in a lookup table. Two distinct weights are simultaneously produced for $(u,v)$ and $(d,p,z)$, respectively. In a possible architecture, the computation of the center of excitation takes 11 clock cycles along with another 10 cycles for fanout. To optimize resources, a single centroid calculation unit can serve multiple engines. Simulations show that a scheme with a unit serving each group of 12 engines is adequately sized for the hit occupancies expected. The search for the local maximum and the center of excitation use local copies of the accumulators so that the incoming hit flux is never stopped unless large fluctuations of the EndEvent word arrival times occur. In these cases the incoming hits are kept on hold within the switch tree.

\subsection{Logic placement, simulation and timing}

Detailed simulations were performed to optimize the physical implementation of our algorithm in a realistic FPGA architecture. 
We studied the implementation in the ALTERA's Stratix V, a device intended for high-bandwidth applications, with optical link inputs, model 5SGXEA7N2F45C2ES.
We programmed all parts of the system in VHDL language, and performed placement and simulation using Altera's proprietary software tools (Quartus II). Each hit was defined as a 41 bits-wide word encoding the corresponding geometric $x$ and $y$ coordinates, a layer identifier, and the associated timestamp. 

In order to serve the needs of a large system, the switching logic was divided in two layers, a pre-switch section to be implemented in a layer of receiving/formatting FPGAs, assumed based on Stratix-V as well, and a second layer containing all the remaning parts of the system, the two being connected by optical fiber links.
The pre-switch section occupies 3.3\% of the available logic in the front-layer Stratix V,  while the remaning part of the switch needs 7.5\% of the logic of the main layer.

Our goal was to fit as many processing engines (the most silicon-demanding part) as possible in each chip. With our baseline design of the processing engines, assuming time-ordered input data, shows that 10$^3$ engines can be fit in one Stratix-V, still leaving enough logic available for the remaining functions. Therefore, our system can fit on order of 100 chips (not accounting for the small logic occupancy required on the input layers, that are assumed to be necessary for other reasons already).

The design of each part has been fully validated through a Modelsim simulation. 
The results show that the whole system is capable of running at a frequencies in excess of 350MHz. The throughput of the switch and of the centroid calculation logic are larger than the processing engines, that thus represents the limiting factor to the overall system throughput.
When accounting for the occupancy predicted by the LHCb simulation in the nominal luminosity conditions of the 2020 upgrade: $L = 2\times10^{33}$ cm$^{-2}$s$^{-1}$ (average of 7.6 interactions per crossing) each engine is loaded with little more than 1 hits/event. Keeping into account the 7 cycles needed to process each hit, this means that the sistem is capable of keeping up with an input frequency of a full 40 MHz of events, already with FPGA currtenly on the market.

The latency budget is shown in Table~\ref{tab:timing}. With clock frequencies of 350 MHz, the total processing latency for reconstructing online tracks is less than  0.5 $\mu$s, thus negligible compared to typical DAQ latencies. This makes the response of the device effectively {\it immediate}, thus making tracks available right after the tracking detectors have been read out.

\begin{table}[tbp]
\centering
\begin{tabular}{lc}
\hline\hline
Task 					& 	Latency (cycles)	\\
\hline
Switching network			&	21				\\
Engine processing			&	70				\\
Clustering				&	55				\\
Data output				&	10				\\
\hline
Total						&	$\simeq 150$			\\
\hline \hline
  \end{tabular}
  \caption{Latency breakdown.}
  \label{tab:timing}
\end{table}

\section{Conclusions}

We showed that high-quality tracking in large LHC detectors is possible at a 40MHz event rate with sub-$\mu$s latencies, when appropriate parallel algorithms are used in conjuction with current high-end FPGA devices.
 
This opens the interesting possibility of designing high-rate experiments where track reconstruction happen transparently as part of the detector readout.


\end{document}